\documentclass{aa}
\usepackage{graphicx}
\usepackage{natbib}
\bibpunct{(}{)}{;}{a}{}{,} 

\begin{document}


\title{HST optical spectral index map of the jet of 3C\,273
\thanks{Based on observations made with the NASA/ESA Hubble Space
Telescope, obtained at the Space Telescope Science Institute, which is
operated by the Association of Universities for Research in Astronomy,
Inc.\ under NASA contract No.~NAS5-26555.} }

\author{S.~Jester \inst{1}
   \and H.-J.~R\"oser \inst{1}
   \and K.~Meisenheimer \inst{1}
   \and R.~Perley \inst{2}
   \and R.~Conway \inst{3}
}

\offprints{jester@mpia.de}

\institute{
   Max-Planck-Institut f\"ur Astronomie, K\"onigstuhl~17,
   69117~Heidelberg, Germany
   \and NRAO, P.\,O.~Box 0, Socorro, NM~87801, USA
   \and Jodrell Bank Observatory, Macclesfield, Cheshire, SK11~9DL, UK
}

\date{Received April 1, 2001/accepted \ldots}

\abstract{We present HST images at 622\,nm and 300\,nm of the jet in
3C\,273 and determine the run of the optical spectral index at
0\farcs2 along the jet.  The smoothness of spectral index changes
shows that the physical conditions are varying smoothly across the
jet.  There is no correlation between the optical flux and spectral
index, as would be expected for relativistic electrons suffering
strong cooling due to synchrotron emission.  We find no evidence for
localized acceleration or loss sites. This suggests that the spectral
shape is not changing much throughout the jet. We show that
relativistic beaming and/or sub-equipartition magnetic fields cannot
remove the discrepancy between light-travel time along the jet and the
lifetime of electrons emitting optical synchrotron radiation.  We
consider this further evidence in favour of a distributed electron
acceleration process.  \keywords{Galaxies: jets -- quasars,
individual: 3C\,273} }

\maketitle


\section{Introduction}
\label{s:intro} While radio jets are a common feature of radio
galaxies and quasars, optical emission has to date been observed
from only about 15 extragalactic jets.  As shown by polarimetric
observations (starting with \citet{Baa56} for M87), both the radio
and optical emission is synchrotron continuum radiation. While
information on the source's magnetic field structure may be
obtained from the polarisation structure, the diagnostic tool for
the radiating particles is a study of the synchrotron continuum
over as broad a range of frequencies as possible, \emph{i.\,e.,} from radio
to UV or even X-ray wavelengths, and with sufficient resolution to
discern morphological details.

The radio and optical emission observed from \emph{hot spots} in
radio jets can be well explained by first-order Fermi acceleration
at a strong shock in the jet (the bow shock)
\citep{MH86,HM87,magnumopus89,hs_II}. But it is not clear that the
optical synchrotron emission from the jet \emph{body}, extending
over tens of kiloparsecs in some cases, can be equally well
explained by acceleration at strong shocks inside the jet.  As is
well known from standard synchrotron theory, electrons with the
highly relativistic energies required for the emission of
high-energy (optical and UV) synchrotron radiation have a very
short lifetime which is much less than the light-travel time down
the jet body in, \emph{e.\,g.}, \object{3C\,273}.  Observations of
optical synchrotron emission from such jets \citep{RM91,M87} as
well as from the ``filament'' near Pictor~A's hot spot
\citep{RM87,PRM97} suggest that both an extended, ``jet-like'' and
a localized, ``shock-like'' acceleration process are at work in
these objects in general and 3C\,273's jet in particular
\citep{hs_II}.  The extended mechanism may also be at work in the
lobes of radio galaxies, where the observed maximum particle
energies are above the values implied by the losses within the hot
spots \citep{Mei96} and by the dynamical ages of the lobes
\citep{BR00}.

The fundamental question is thus: how can we explain
high-frequency synchrotron emission far from obvious acceleration
sites in extragalactic jets?  Although most of the known optical
jets are very small and faint \citep{SU00}, there are a few jets
with sufficient angular size and surface brightness to be studied
in detail: those in M87 (a radio galaxy), PKS~$0521-365$ (an
elliptical galaxy with a BL\,Lac core), and 3C\,273 (a quasar).

We have embarked on a detailed study of the jet in 3C\,273 using
broad-band observations at various wavelengths obtained with
today's best observatories in terms of resolution: the VLA (in
combination with MERLIN data at $\lambda$6\,cm) and the HST. Using
these observations, we will derive spatially resolved (at
0\farcs2) synchrotron spectra for the jet.  3C\,273's radio jet
extends continuously from the quasar out to a terminal hot spot at
21\farcs5 from the core, while optical emission has been observed
only from 10\arcsec\ outwards.\footnote{For the conversion of
angular to physical scales, we assume a flat cosmology with
$\Omega_{\mathrm{m}}=0.3$ and $H_{0} = h_{60}\times
60\,\mathrm{km\,s^{-1}Mpc^{-1}}$, leading to a scale of $3.2
h_{60}^{-1} \mathrm{kpc}$ per second of arc at 3C\,273's redshift
of 0.158.} On ground-based images, the optical jet appears to
consist of a series of bright knots with fainter emission
connecting them.  So far, synchrotron spectra have been derived
for the hot spot and the brightest knots using ground-based
imaging in the radio \citep{jetII}, near-infrared
$K^{\prime}$-band \citep{NMR97} and optical $I, R, B$-bands
\citep{RM91} at a common resolution of 1\farcs3
\citep{MNR96,Roe00}. The radio-to-optical continuum can be
explained by a single power-law electron population leading to a
constant radio spectral index\footnote{We define the spectral
index $\alpha$ such that $f_\nu \propto \nu^{\alpha}$.} of $-0.8$,
but with a high-energy cutoff frequency decreasing from
$10^{17}\,$Hz to $10^{15}$\,Hz outwards along the jet.  The aim of
the study is both the determination of the spectral shape of the
synchrotron emission, and by fitting synchrotron spectra according
to \citet{magnumopus89}, deriving the maximum particle energy
everywhere in the jet.  The \emph{observed} spectra can then be
compared to \emph{predictions} from theoretical work.
\begin{table*}
\begin{center}
\begin{tabular}{llllllll}
\hline\hline
Filter & Mean $\lambda$ & FWHM & Exposure & \multicolumn{2}{l}{Point
source} & \multicolumn{2}{l}{Extended source}\\
       & nm             & nm   & s    & mag      & $\mu$Jy  & mag/$\sq \arcsec$ & $\mu$Jy/$\sq \arcsec$ \\
\hline
F300W  & 301 & 77 & 35\,500 &  26.1 & 0.04     & 20.8 & 5.2 \\
F622W  & 620 & 92 & 10\,000 &  27.7 & 0.04     & 23.0 & 2.8 \\
\hline \hline
\end{tabular}
\end{center}
\caption{Passbands and limiting magnitudes for the observations. Point
source: $10\sigma$ detection limit, prediction by the exposure time
calculator. Extended source: $5\sigma$ per pixel detection limit,
determined from the background noise measured on reduced
frames. Magnitudes are Vegamags referred to the corresponding HST
filter band.}
\label{t:lim_mag}
\end{table*}

As an intermediate result of our study, we present HST WFPC2 images of
the jet in 3C\,273 in a red (F622W) and near-UV (F300W) broadband
filter. The near-UV observations constitute the highest-frequency
detection of synchrotron emission from 3C\,273 so far.  (Extended X-ray
emission has also been observed, but it is unclear at present whether
this, too, is due to synchrotron radiation
\citep{Roe00,Mareta01,Sam01}.)  From these images, we construct an
optical spectral index map at 0\farcs2 resolution.

After a description of the observations and the data reduction in
Sections \ref{s:observations} and \ref{s:data}, we examine the direct
images in Sect.~\ref{s:image}.  The creation and description of the
spectral index map follow in Sect.~\ref{s:index}.  We analyse the map
in Sect.~\ref{s:discussion} and conclude in Sect.~\ref{s:conc}.  Details
regarding the alignment of HST images are found in
App.~\ref{a:alignment}.


\section{Observations}
\label{s:observations}

Observations were made using the Planetary Camera of the second
Wide Field and Planetary Camera (WFPC2) \citep{wfpc} on board the
{\sc Hubble Space Telescope}\footnote{Proposal ID 5980} in three
sets of observations
on March 23rd and June 5th/6th, 1995, for 35\,500\,s
through filters F300W (ultra-violet, $U$) and
10\,000\,s F622W (roughly $R_{\mathrm{C}}$ in the Kron-Cousins
system).  The resulting limiting magnitudes are listed in
Table~\ref{t:lim_mag}.  The jet was imaged onto the center of the
Planetary Camera (PC) chip which has 800 by 800 pixels with a
nominal size of 0\farcs0455 projected on the sky. The telescope
was oriented such that the position angle of the PC chip's y-axis
was $\approx 154\degr$. This way, the quasar itself is also mapped
on the PC, while each of three neighbouring astrometric reference
stars \citep[see Table~3 in ][]{RM91} was observed on one of the three
Wide Field chips. This choice gives distortion-free images of the
jet, while allowing the use of the quasar and the reference stars
for image alignment.

The total exposure time was split into individual exposures of about
2500\,s or one HST orbit, executed in three separate sets of observations. Each of
these exposures was done at a telescope pointing slightly offset from
all others (by up to 1\arcsec) to facilitate the correction for chip
artifacts. One short exposure was obtained in each set of observations and filter to
measure the positions of the quasar and reference stars, most of which
are saturated on the long exposures.
Our study of 3C\,273's jet will be conducted employing a 0\farcs2
effective beam size.  This results in a typical signal to noise ratio
($S/N$) \emph{per resolution element} in the red band of around 100 in the brightest
regions and 30--40 in inter-knot regions. Because both
the jet flux and the WFPC2 and telescope throughput decrease towards
the UV, the UV-band $S/N$ is only about 40 in the brightest regions,
while the inter-knot regions are barely detected.


\section{Data reduction}
\label{s:data}

The data were re-calibrated under \verb+IRAF+ with the \verb+STSDAS+
package provided by STScI, using the ``most recent'' calibration
reference files as of December 1998, because some of the files had
changed since the original observations. The values of background
noise measured on the calibrated frames agree well with the values
expected from photon statistics, as calculated from the expected read
noise, dark current and sky background level.

One of the PC chip's charge traps \citep{ISR-CT,Data} lies inside
the jet image, in column 339. This has no observable effect on the
faint UV image, but the effect had to be corrected on the
well-exposed red-band images. This was done by replacing the
affected portion of each image by the corresponding pixels from an
offset image, as a correction according to \citet{ISR-CT} unduly
increased the noise in the corrected part of the image.

The images were initially registered using the commanded offsets to
the nearest pixels.  This alignment is sufficient for the rejection of
cosmic rays as these only affect a small number of adjacent pixels.
Cosmic rays were rejected using a standard $\kappa$-$\sigma$ algorithm,
rejecting all pixel values deviating more than $4 \sigma$ from the
local (low-biased) median in a first pass, and neighbouring pixels
with more than $2.5 \sigma$ deviation in a second pass. The number of
pixels treated this way agrees with the expected cosmic ray hit rate
for the images.

A model of the sky background and ``horizontal smear'' \citep[increased
pixel values in rows containing saturated pixels from the quasar's
core, Chap. 4 of][]{wfpc} was fitted in the part of the image containing the jet using
second-order polynomials along rows. The coefficients of the
polynomials were then smoothed in the perpendicular direction. The
conversion from photon count rate to physical flux units used the
throughput information provided by STScI which is valid after the 1997
\verb+SYNPHOT+ update.
\begin{figure*}
  \sidecaption
  \includegraphics[width=12cm]{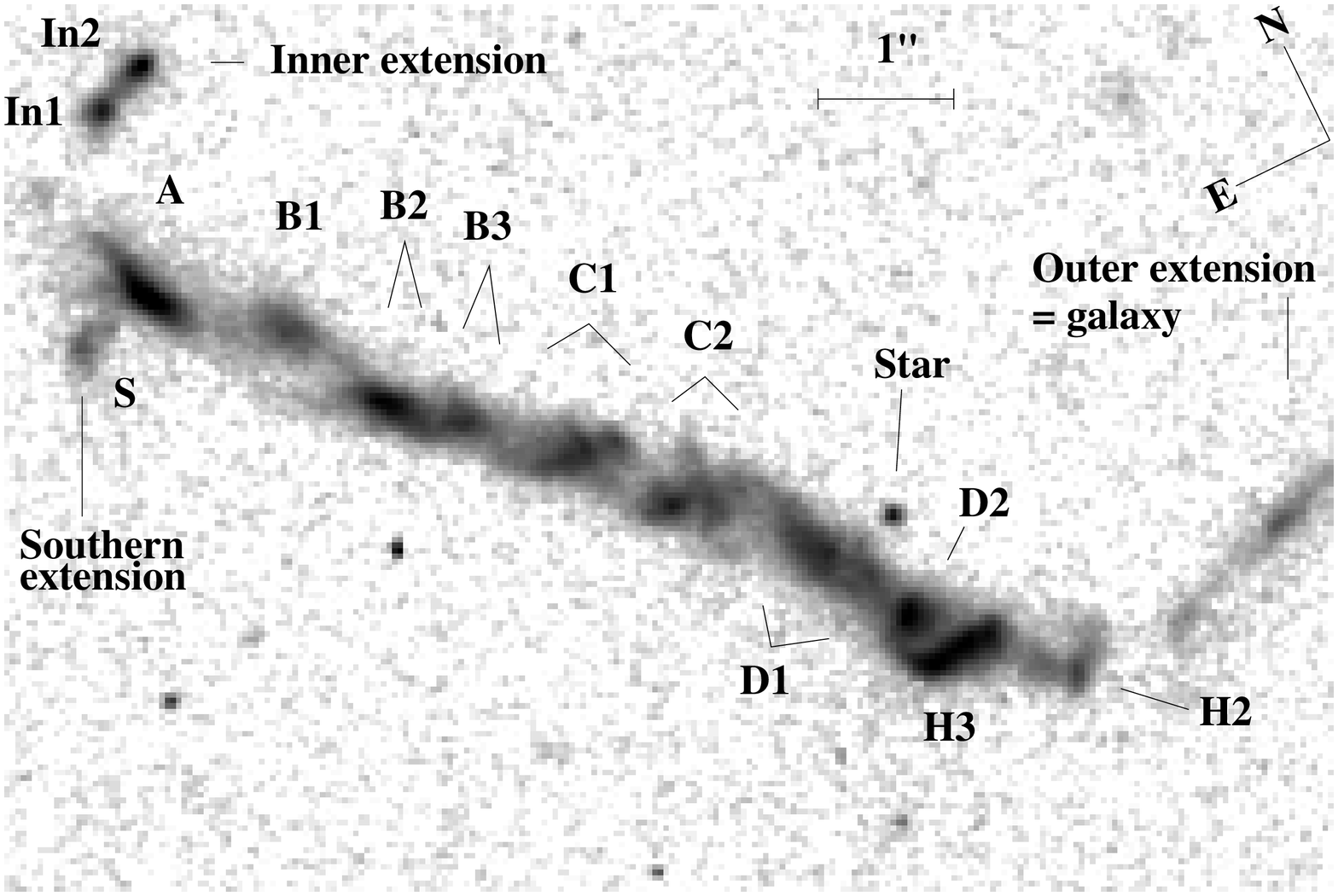}
  \caption{The jet in red light (620\,nm) after background
  subtraction. Logarithmic grey-levels run from 0 to
  0.04$\mu$Jy/pixel,  0\farcs08 effective beam size, 0\farcs045 pixel
  size. The quasar core lies 10\arcsec\ to the northeast
  from A.  The labelling of the jet features as introduced by
  \citet{Leleta84} and extended by \citet{RM91}, together with the hot spot nomenclature
  from \citet{FC85} is also shown. Note that the
  labelling used by \citet{Baheta95} is slightly
  different.}
  \label{f:R-band}
\end{figure*}
\begin{figure*}
  \resizebox{12cm}{!}{\includegraphics{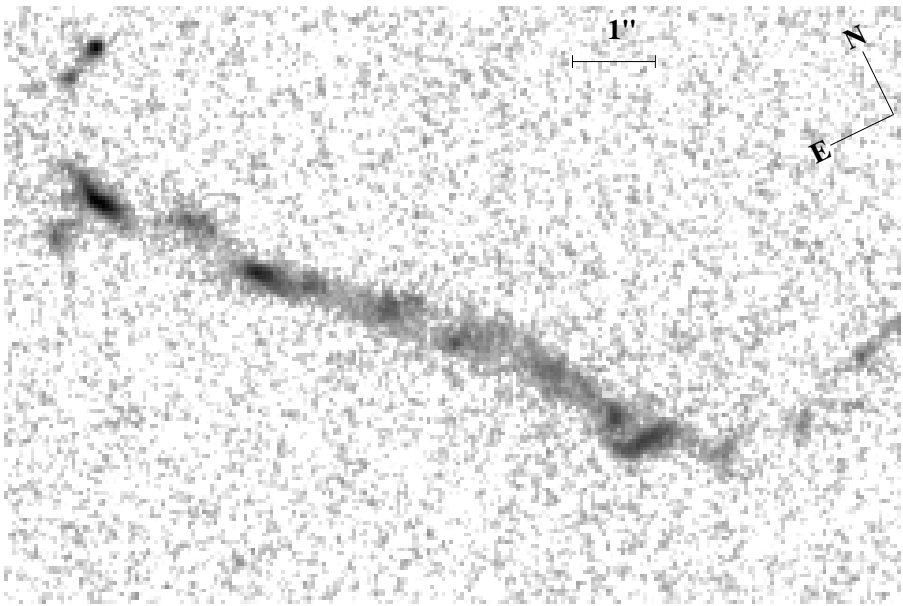}}
  \hfill
  \parbox[b]{55mm}{
    \caption{The jet in UV light (300\,nm) after background
    subtraction. Logarithmic grey-levels run from 0 to
  0.014$\mu$Jy/pixel, 0\farcs06 effective beam size, 0\farcs045 pixel size.}
    \label{f:U-band}
  }
\end{figure*}

\section{Maps of the optical brightness}
\label{s:image} The calibrated images are presented in
Figs.~\ref{f:R-band} and \ref{f:U-band}. The morphology of the jet is
identical in both images and appears rather similar to the morphology in
high-resolution radio maps \citep{jetII,Baheta95}.  The exception to
this is the radio hot spot, being the dominant part in the radio but
fairly faint at high frequencies. Our images show structural details
of the optical jet which were not discernible on earlier, shallower
and undersampled HST WF images of 0\farcs1 pixel size
\citep{Baheta95}. Based on our new maps, the term ``knots'' seems
inappropriate for the brightness enhancements inside the jet, as these
regions are resolved into filaments.  The higher resolution
necessitates a new nomenclature for the jet features
(Fig.~\ref{f:R-band}).  For consistency with earlier work
\citep{Leleta84,FC85,RM91}, our nomenclature is partly at variance
with that introduced by \citet{Baheta95}.

The jet is extremely well collimated -- region A has an extent
(width at half the maximum intensity) of no more than 0\farcs8
perpendicular to the average jet position angle of $\sim 222\degr$
(opening angle $\la 5\degr$). Even region D2/H3 is only
1\arcsec\ wide (opening angle $\approx 2.5\degr$). The optical jet
appears to narrow towards the hot spot, in the transition from H3
to H2.

Region~A is now seen to extend further towards the core than
previously known.  It may be noteworthy that \citet{Leleta84}
reported the detection of an extension of knot A towards the
quasar, whose existence at the reported flux level was not,
however, confirmed by later work.

The criss-cross pattern visible in regions C1 and C2, and less
clearly in B1-2 and D1, is reminiscent of a (double?)  helical
structure \citep{Baheta95}, but could also be explained by oblique
double shocks \citep{HN89}.

The jet has three ``extensions'' (Fig.\,\ref{f:R-band}), none of
which has been detected at radio wavelengths.  The morphology of
the \emph{outer extension} supports the classification as a galaxy
based on its colours made by \citet{RM91}.  The nature of the
other two extensions, however, remains unknown even with these
deeper, higher resolution images.  The \emph{northern inner
extension} was already resolved into two knots (In1, In2) on a
{\sc Faint Object Camera} image \citep{TMW93}.  The two knots are
extended sources and clearly connected to each other.  The
\emph{southern extension} is featureless and an extended source.

Comparing the direct images, we can immediately estimate that the jet's
colour slowly turns redder outwards from region~A.  The similarity of
the jet images in both filters shows that there are no abrupt colour
changes within the jet.  For a quantitative assessment of the
extensions' and the jet's colour in the following, we derive an
optical spectral index map.


\section{Optical spectral index map}
\label{s:index}

Firstly, we consider how to derive a map of the spectral index
$\alpha_{RU}$ from the presented images, at a common effective beam
size of 0\farcs2.
\begin{figure*}
  \resizebox{12cm}{!}{\includegraphics{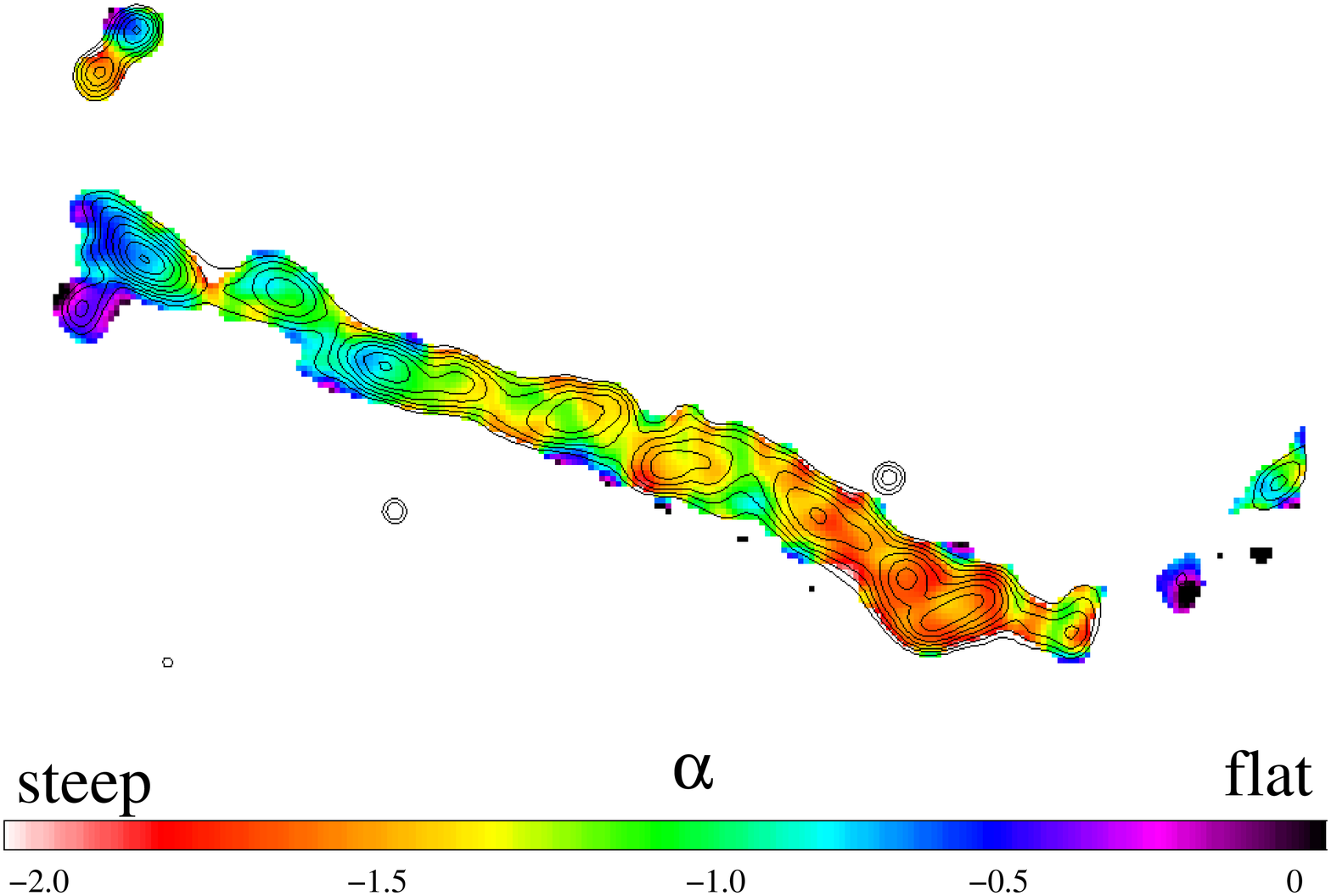}}
  \hfill
  \parbox[b]{55mm}{
    \caption{Optical spectral index $\alpha_{RU}$ ($f(\nu) \propto
    \nu^{\alpha}$) at 0\farcs2 resolution. The contours show the
    red-band image and are logarithmic
    with a factor $\sqrt{2}$ from 30 $\mu$Jy/beam to 340
    $\mu$Jy/beam.}
    \label{f:alpha}
  }
\end{figure*}

\subsection{Definition of optical spectral index}
\label{s:index.definition}

Independently of whether a spectrum actually does follow a power law
over any range of frequencies, a local two-point spectral index can be
defined between any two surface brightness measurements $B_1, B_2$ at
frequencies $\nu_1$ and $\nu_2$ { (with $\nu_2 < \nu_1$)},
respectively, as
\begin{equation}
\alpha = \frac{\ln\frac{B_1}{B_2}}{\ln \frac{\nu_1}{\nu_2}}.
\label{eq:alpha}
\end{equation}

The error on the spectral index is computed from the values of the
noise $\sigma_1, \sigma_2$ in the respective input images:
\begin{equation}
\sigma_{\alpha} = \frac{1}{\ln\frac{\nu_1}{\nu_2}}
\sqrt{\frac{\sigma^{2}_1}{B^{2}_1} +
\frac{\sigma^{2}_2}{B^{2}_2}}.
\label{eq:sigma_alpha}
\end{equation}
This formula shows that the spectral index error depends on the $S/N$
of the input images and the ``baseline'' between the wavelengths at
which the observations are made. In the following, we will use $\nu_1
= \nu(U)$ for the UV and $\nu_2 = \nu(R)$ for the red, together with
$B_1 = B_U, B_2 = B_R$. With the mean wavelengths of the filters,
2942.8\,\AA\ for F300W and 6162.8\,\AA\ for F622W \citep{wfpc}, a cut
on the signal-to-noise ratio $S/N > S/N_{\mathrm{cut}}$ of the input
images limits the spectral index error to at most
\begin{equation}
\sigma_{\alpha_{RU}} \leq { \frac{1}{\ln \frac{6162.8}{2942.8}}}
\sqrt{\frac{2}{S/N_{\mathrm{min}}}} = \frac{1.9}{S/N_{\mathrm{min}}}.
\label{eq:sn-cut}
\end{equation}
In our case, the UV-band $S/N$ is always much inferior to that in the
red band, so the error will actually be dominated by the UV noise and
hence smaller than the maximum from Eqn.~\ref{eq:sn-cut} for the most
part.  The noise in the input images is calculated from the calibrated
images and includes shot noise due to both observed photons and dark
current as well as the read noise, but excludes systematic errors,
whose expected magnitude we now examine.

\subsection{Systematic errors for a spectral index determination}
\label{s:index.errors}

We consider the systematic errors which may be introduced when
combining two images taken through different filters, or by different
instruments and telescopes.  The main danger in a determination of
spectral gradients lies in a misalignment between the images which
would introduce spurious gradients; referring the flux to different
effective beam sizes will lead to wrong spectral index determinations
as well.  Following considerations given in Sect.~\ref{s:flux_error},
we deduce that a 5\% limit on the flux error due to misalignment
requires aligning the images to better than 10\% of the effective PSF
full width.  The error in the PSF determination is negligible when, as
in our case, the smoothing Gaussian is much wider than the PSF.

The 0\farcs2 effective resolution aimed for thus requires knowing the
\emph{relative} alignment of all images in the data set to better than
20\,mas or 0.44 PC pixels.  The \emph{absolute} telescope pointing
does not need to be known for this purpose as we tie all positions
to the quasar core as origin.

After detailed investigations of all issues related to relative
alignment of HST images (see Sect.~\ref{s:alignment}), we used the
following procedure: as the first step, all exposures through one
filter and within one set of observations are summed up using the
relative positions from the ``jitter files'' provided as part of
the observing data package (typical error 5~mas).  The exposures
through F300W were distributed over three different sets of
observations, so we obtain three intermediate images.  For
aligning these with each other, we use the average of the
positional shifts of the quasar and three astrometric stars
(observed on the Wide Field chips) measured on the intermediate
image or the short exposure in each set of observations (error
10~mas).  The field flattener windows inside WFPC2 introduce a
wavelength dependence of the plate scale, which has to be removed
prior to combination of the images taken through different
filters.  We therefore sample all four intermediate images (three
UV and one red) onto a grid of one tenth of their average plate
scale with a pixel size of 0\farcs0045548.  We sum the UV
intermediate images to give a UV final image.  Because of the
scale change, we have to align this to the red image using the
quasar position only (estimated error 10~mas--15~mas).  We then
rebin both images to the final pixel size of 0\farcs045548. Adding
up all errors in quadrature, the error margin of 20~mas is just
kept.

Note that the rebinning has only matched the scales of the two images,
but not removed the geometric distortion of the focal plane.  The
next-order wavelength-dependent term in the geometric distortion
solution is at least two orders of magnitudes smaller than the scale
difference.  Both the red and the ultraviolet image can thus be
assumed to have identical distortion solutions.  We therefore ignore
the geometric distortion for the remainder of this paper, but caution
that it must be taken into account when comparing these data with
data obtained at other instruments.

\subsection{Calculation of the optical spectral index map}
\label{s:index.map} We have now aligned the images in the two
filters to better than 0\farcs02. When comparing flux measurements
on these images, they have to be referred to a common beam size.
The maps are therefore smoothed to a common 0\farcs2 using
Gaussians with a width matched to the effective FWHM of the PSF in
each image. This operation retains the original pixel size.  As
few point sources are available, the effective FWHM is determined
on PSF models generated using the \verb+TinyTIM+
software~\citep{tinytim}, resulting in 0\farcs08 on the red and
0\farcs06 on the UV-band image. Because of sampling effects, the
measured widths are larger than the size of the Airy disk of the
HST's primary at the corresponding wavelengths.

\begin{figure*}
  \resizebox{12cm}{!}{\includegraphics{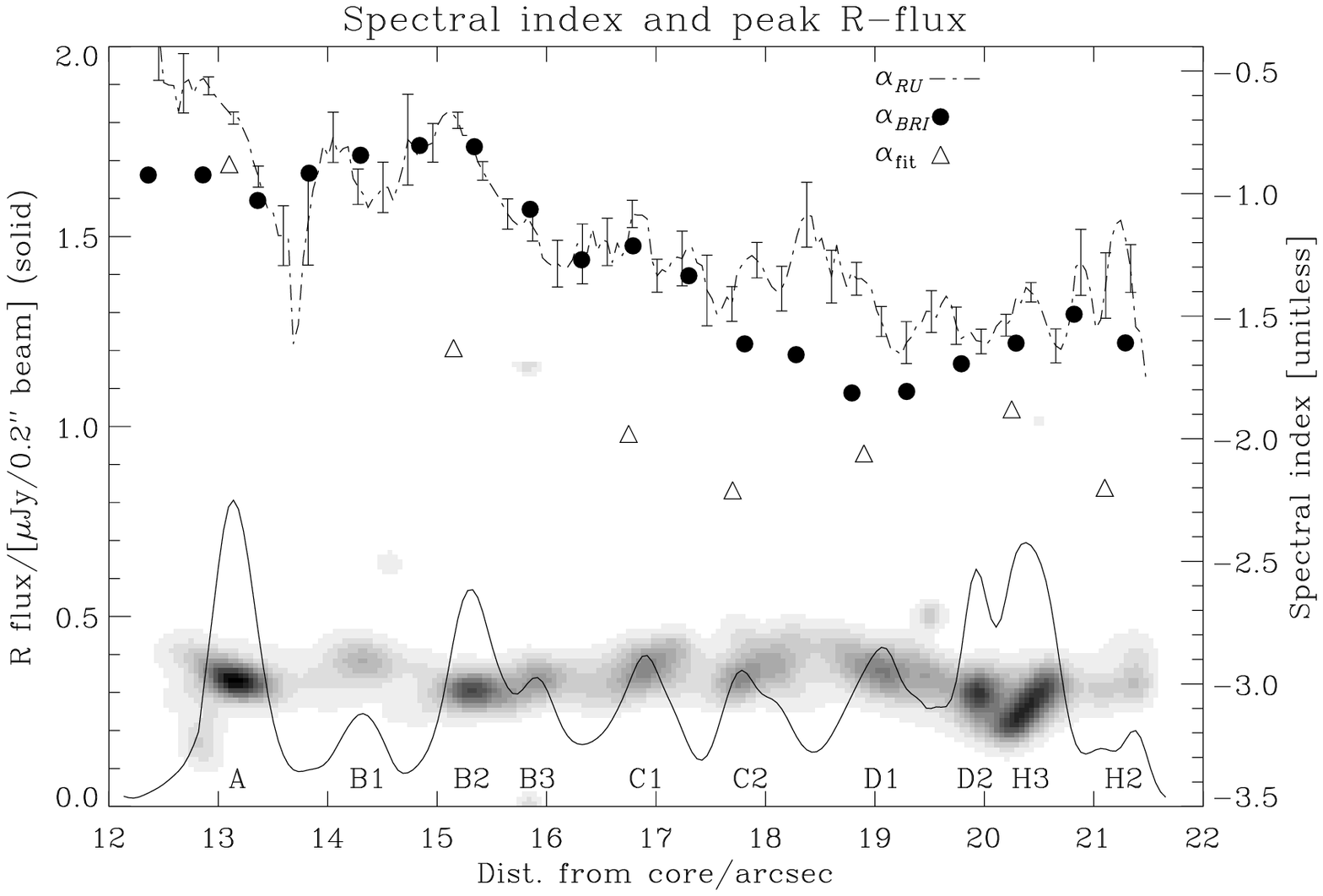}}
  \hfill
   \parbox[b]{55mm}{
     \caption{Run of the red-band brightness and optical spectral index
     along the outer half of the jet in 3C\,273, for a 0\farcs2
     beam. $\alpha_{RU}$ was determined from Figures 1 and 2, while
     $\alpha_{BRI}$ for a 1\farcs3 beam is taken from
     \citet{RM91}. For comparison, we show $\alpha_{\mathrm{fit}}$,
  the corresponding spectral
     index obtained from synchrotron
  spectra fitted by \citet{MNR96} and \citet{Roe00}.  While the
  observed spectral index agrees with older data, it is now clear that
  the fit is inadequate for the optical part of the spectrum.  The
  steeper spectral index of the fit may either be
  due to contamination of the infrared flux by a ``backflow'' component
  around the jet, or because of the presence of a second high-energy electron
  population in the jet which is not included in the fit (see
  Sect.~\ref{s:disc.jet.alpha} for a discussion of the discrepancies).
     \label{f:alpha-trace}
   }
  }
\end{figure*}

Fig.~\ref{f:alpha} shows the spectral index map obtained from the
images (Figs.~\ref{f:R-band} and \ref{f:U-band}) according to
Eqn.~\ref{eq:alpha}.  Only those points are shown in the map which
have an aperture $S/N$ of at least 5 on both images.  This limits
the error in the spectral index to 0.4 (Eqn.~\ref{eq:sn-cut}). The
error is less than 0.1 inside all knot regions, so that a colour
change indicates statistically significant variations of the
spectral index. Fig.~\ref{f:alpha-trace} shows the spectral index
along a tracing of the jet obtained from a rotated map with the
jet's mean position angle of 222\fdg2 along the horizontal.

\subsection{Colours of the extensions}
\label{s:index.extensions} The \emph{outer extension} is not
bright enough to show up on the map in its entirety.  The
variation of $\alpha_{RU}$ across it are consistent with a spiral
galaxy.  The \emph{inner extension's} two knots have markedly
different colours on the spectral index map.  Knot In2 shows a
spectral index gradient, $\alpha \approx -0.9$ to $-0.3$ roughly
parallel to the jet and outwards from the quasar position.  Knot
In1 has a much steeper spectral index of about $-1.5$ and shows no
gradient.  Interestingly, the \emph{southern extension} S has the
flattest spectrum ($\alpha_{RU} \approx -0.4$) of all regions on
the map.

\subsection{Spectral index of the jet}
\label{s:index.jet}

The optical spectral index declines globally outwards from $-0.5$ near
the onset of the optical jet at A to $-1.6$ in D2/H3.  This trend does
not continue into the hot spot, again stressing the physical
distinction between jet and hot spot.  The general steepening is in
agreement with previous determinations of the knots' synchrotron
spectrum which showed a decline of the cutoff frequency outwards
\citep{MNR96,Roe00}.  The optical spectral index $\alpha_{BRI}$
determined at 1\farcs3 resolution \citep{RM91} agrees very well with
our new determination of $\alpha_{RU}$ at the much higher resolution
(Fig.~\ref{f:alpha-trace}).  { The only discrepancy arises in C2/D1;
we defer a discussion to Sect.~\ref{s:disc.jet.alpha}.} The smooth
variations of the spectral index along the jet show that the physical
conditions in the jet change remarkably smoothly over scales of many
kpc.  There is no strict correlation between red-band surface
brightness and spectral index like that found in the jet in M87
\citep{MNR96,Per01}.

There is a marginally significant flattening of the spectrum in the
transitions A-B1, B1-B2, C1-C2, and moving out of C2, consistent with
the \emph{absence of losses}.  In any case, the overall steepening of
the spectrum (from region~A down do D2/H3) is less rapid than that
\emph{within} individual regions (\emph{e.\,g.}, A and B2).  There is no
significant steepening from D1 out to the bridge between H3 and H2,
despite large variations in surface brightness.

The criss-cross morphology in C1, C2, B1, B2 and D1 is reflected on
the spectral index map as a band of one colour crossing a second one.
This is most clearly seen in knot C1 which has a green band of $\alpha
\approx -1.1$ across an orange region of $\alpha \approx -1.4$,
supporting the interpretation of two emission regions appearing on top
of each other.  The spectral index near the hot spot shows a flip from
flat ($-1.0$) to steep ($-1.5$).


\section{Discussion}
\label{s:discussion}

\subsection{The jet's extensions}
\label{s:disc.ext}
\begin{figure}
\center \resizebox{0.8\hsize}{!}{\includegraphics{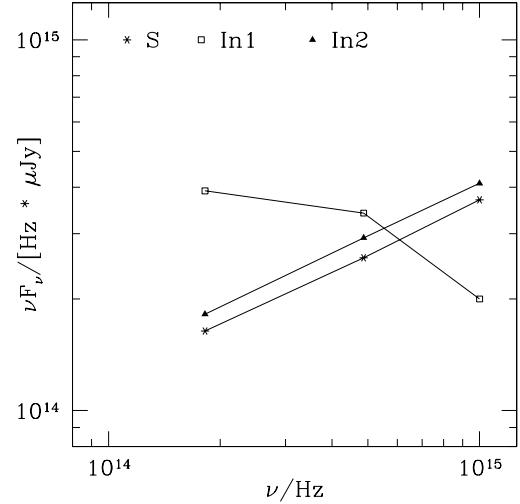}}
  \caption{Broad-band flux points for the radio-quiet extensions to
  the jet from HST imaging at 1.6\,$\mu$, 620\,nm and 300\,nm.\label{f:extensions}}
\end{figure}

Before analysing the jet emission proper, we consider the relation of
the extensions to the jet.  Our new images confirm the \emph{outer
extension's} nature as spiral galaxy.  The spectral index map gives us
few hints about the nature of the emission mechanism for the
\emph{inner} and {southern extensions}.  We have therefore measured
their \emph{integral fluxes} at 300\,nm and 620\,nm (this paper) and
at 1.6$\mu$ (HST NICMOS camera~2 imaging\footnote{Proposal ID 7848},
\emph{in prep.}).  Fig.~\ref{f:extensions} shows the resulting
spectral energy distributions.  In1, with the steepest $\alpha_{RU}$,
is also the brightest in the infrared.  The infrared flux points of
both extension S and of In2 lie on the continuation of the power law
with index $-0.4$, determined on the optical spectral index map.  If
this power law were valid up to radio wavelengths, it would predict a
flux at $\lambda 6$\,cm of about 50$\mu$Jy for the northwestern knot
In1, and of about 45$\mu$Jy for the southern extension S.  This is a
factor of order 100 below the flux from the jet at this wavelength
\citep{jetII}, and any emission at this level would not be detected
even on our new VLA maps with an RMS noise of $\approx 0.8$\,mJy.
There is no similarly obvious extrapolation of In1's SED to radio
wavelengths.  A full comparison will be presented in a future paper.

The inner extension shows a polarisation signal on unpublished
ground-based maps made by our group and on polarimetric images
obtained with the Faint Object Camera on board HST \citep{TMW93}.  A
polarisation signal could be due to scattered quasar light
\citep{RM91} or, together with possible radio emission, synchrotron
radiation.  We aim to clarify the nature of the inner extensions by
planned polarimetric and spectroscopic observations with the VLT.

\subsection{The hot spot}
\label{s:disc.hotspot} The optical counterpart to the radio hot
spot (H2) appears very faint on both images, and the spectral
index map shows a flip from flat to steep there.  Both features
are explained by the \citet{MH86} hot spot model: the lifetime of
electrons emitting in the optical is quenched by the strong
magnetic field in the hot spot. The spectral index flip is
expected if there is an offset between the emission peaks at
different frequencies.  This offset is predicted by the theory by
\citet{MH86} and is clearly seen when comparing optical and radio
images (\citet{Roeeta97}; a detailed comparison of all data will
be presented in a future paper).  By its spectrum and morphology,
the hot spot appears exactly as expected for a high-loss
synchrotron emission region downstream of a localized strong shock
with first-order Fermi acceleration.  Finally, we note that
although the term ``hot spot'' is historically in use for this
part of the jet, the term is ill-defined and may have become
inadequate, but we defer a discussion of its adequacy to a future
paper.

\subsection{Main body of the jet}
\label{s:disc.jet}

Our HST images (Figs.~\ref{f:R-band} and \ref{f:U-band}) show a close
coincidence of the jet's morphology over a factor of 2 in frequency
(optical and UV).  The spectral index map (Fig.~\ref{f:alpha}) shows
amazingly smooth variations in the physical conditions over the entire
jet.  The electrons' synchrotron cooling leaves little imprint on the
spectral index along the jet, contrary to expectations.

\subsubsection{The spectral index at HST resolution}
\label{s:disc.jet.alpha}
\begin{figure}
 \resizebox{\hsize}{!}{\includegraphics{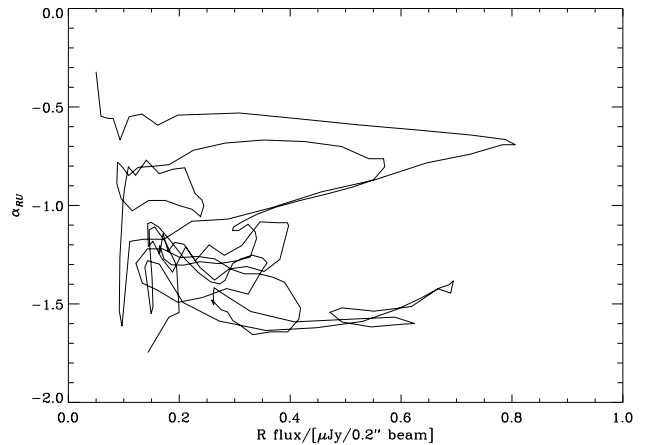}}
 \caption{Optical spectral index against red-band brightness for the
 outer half of the jet in 3C\,273 (data as in Fig.~\ref{f:alpha-trace}).
 There is no strict correlation like that observed in the jet of M87.}
 \label{f:correlation}
\end{figure}
The two striking features of our spectral index map are the smooth
variation of the spectral index over the entire jet
(Fig.~\ref{f:alpha-trace}), and the lack of a strong correlation
between red-band brightness and spectral index
(Fig.~\ref{f:correlation}) like that observed in the jet of M87
\citep{MNR96,MRS96,Per01}. There are large but smooth variations
of the spectral index, while the surface brightness remains fairly
constant over the jet's projected extent of about 10\arcsec.
Conversely, there are large \emph{local} variations of surface
brightness without strong changes in the spectral index.  We do
not observe abrupt changes of the spectral index inside the jet
like we do in the hot spot.  The run of the spectral index is thus
consistent with the complete absence of energy losses over scales
of many kiloparsecs, in spite of the large observed synchrotron
luminosity, which indicates the need for re-acceleration of
particles inside the jet.  The mismatch between synchrotron
cooling scale and extent of the jet even led \citet{GN75} to
conclude that the jet's emission could not be synchrotron
radiation.

The same lack of apparent synchrotron cooling is implied by the
surprising overall correspondence between $\alpha_{BRI}$ and
$\alpha_{RU}$ (Fig.~\ref{f:alpha-trace}) which have been determined at
vastly different resolutions (1\farcs3 and 0\farcs2, corresponding to
$4.2h_{60}$\,kpc and $640h_{60}$\,pc, respectively).  The small
differences (strongest in regions A and C2/D1) can be explained at
least in part by the different beam sizes: the spectral index
determined by smoothing the HST images to 1\farcs3 makes the
discrepancy smaller.  In addition, the emission region becomes wider
towards longer wavelengths, an effect already noticeable in the
near-infrared $K^\prime$ band \citep{NMR97} and attributed to a
``backflow'' \citep{jetIV}. The presence of such a steep-spectrum,
diffuse component more extended than the jet channel might also
explain the discrepancy.

The general agreement indicates a constancy of the shape of the
spectrum even at the scales resolved by HST.  { To first order, the
appearance of the jet at all wavelengths can be explained through such
a constant spectral shape consisting of a low-frequency power law and
a high-frequency curved cutoff.  The assumed spectral shape is shifted
systematically in the $\log\nu-\log S_{\nu}$ plane, passing different
parts of the curved cutoff through the optical flux point.  This leads
to similar morphological features at all wavelengths, in accord with
observations \citep{jetII,Baheta95}.  Moving the spectral shape
through a constant optical flux point leads to a correlation of
steeper optical spectral index with higher radio flux. This
correlation reproduces the observed tenfold increase of the radio
surface brightness from A to D2/H3 \citep{jetII} and the overall
steepening of the optical spectral index.} However, this steepening
happens on much longer time scales than expected from synchrotron
cooling alone.

Assuming an electron energy distribution with a fairly sharp cutoff,
drastic jumps in $\alpha_{RU}$ would be expected at the locations of
the shock fronts if the jet knots were (strong) shocks like the hot
spot without extended re-acceleration acting between them.  The
absence of strong cooling makes it impossible to pinpoint localised
sites at which particles are either exclusively accelerated or
exclusively undergo strong losses.  Any acceleration site must
therefore be considerably smaller than the beam size we used
($640h_{60}^{-1}$\,pc), and these acceleration sites must be
distributed over the entire jet to explain the absence of
cooling. This, together with the low-frequency spectral index of
$\approx -0.8$, corresponds to the jet-like acceleration mechanism
proposed by \citet{hs_II}.

\citet{MNR96} have presented fits of synchrotron continua to the
observed SEDs of knots A, B, C, D and the hot spot H at 1\farcs3
which have also been used by \citet{Roe00}. It is noted that the
optical spectral index predicted from these spectra
($\alpha_{\mathrm{fit}}$ in Fig.~\ref{f:alpha-trace}) is always
steeper than the observed spectral index, that is below
$\alpha_{BRI}$ and $\alpha_{RU}$. This indicates that the fitted
spectrum is not fully adequate at the highest frequencies.  This
may again be due to contamination of the near-infrared flux by the
same ``backflow'' component mentioned above. The contamination
would make the IR-optical spectral index (which dominates the run
of the spectrum at high frequencies) steeper than the optical
spectral index, as observed.  Alternatively or additionally, there
could be deviations of the optical-UV spectral shape from a
standard synchrotron cutoff spectrum which may be interpreted as
the first observational hint towards the existence of a second,
higher-energy electron population producing the X-ray emission
\citep{Roe00}. Detailed statements about the adequacy of model
spectra have to be deferred to a future paper considering the full
radio, infrared and optical data set.

\subsubsection{Can beaming account for the lack of cooling?}
\label{s:disc.jet.cooling}

\citet{HB97} proposed that sub-equipartition magnetic fields
combined with mildly relativistic beaming could explain the lack
of cooling in the jet of M87 -- which is, however, ten times
shorter than that of 3C\,273.  As an alternative to postulating
re-acceleration, we consider whether low magnetic field values and
beaming could lead to electron lifetimes sufficient to allow
electrons to be accelerated at region~A to illuminate the entire
jet down to the hot spot over a projected extent of $32
h_{60}^{-1}$\,kpc (the argument will become even more stringent by
demanding acceleration in the quasar core). We consider the
electron lifetime against synchrotron and inverse Compton cooling
off { cosmic} microwave background photons; the synchrotron
self-Compton process is negligible for electrons in the jet
\citep{Roe00}, as is Compton scattering off the host galaxy's star
light.

The total energy loss rate of an electron with energy $E$ due to
synchrotron radiation and inverse Compton scattering, averaged
over many pitch-angle scattering events during its lifetime, is
\begin{equation}
-\frac{\mathrm{d}E}{\mathrm{d}t} = \frac{4}{3} \sigma_{\mathrm{T}} c
 U_\mathrm{tot} \beta^2 \left(\frac{E}{m_{e}c^2}\right)^2,
\label{eq:eloss}
\end{equation} where $U_\mathrm{tot} = U_{\mathrm{CBR}}(z) +
 U_\mathrm{mag}$ is the sum of the energy densities of the background
radiation and magnetic field, respectively, and
$\sigma_{\mathrm{T}}$ is the Thomson cross-section
\citep{Longair_eloss}.  We integrate this equation from $E=\infty$
at $t=0$ to $E(t)$, assuming $\beta = 1$ (appropriate for the
highly relativistic electrons required for optical synchrotron
radiation) and substitute for the electron's energy $E = \gamma
m_{e}c^2$.  Inverting yields the maximum time that can have
elapsed since an electron was accelerated, given its Lorentz
factor $\gamma$ \citep{vdLP69}:
\begin{equation}
t(\gamma) = \frac{m_{e}c^2}{ \frac{4}{3} \sigma_{\mathrm{T}} c
 U_\mathrm{tot} \gamma}.
\label{eq:tofE}
\end{equation}
This is the ``electron lifetime'', inversely
proportional to both the energy density in which the electron has
been ``ageing'', and the electron's own energy.

As most of the electron's energy is radiated at the synchrotron
characteristic frequency $\propto \gamma^2 B$, we can substitute
for $\gamma$ in Eqn.~\ref{eq:tofE} in terms of the observing
frequency and the magnetic field in the source.  Hence,
Eqn.~\ref{eq:tofE} becomes (in convenient units) \begin{equation}
t_{\mathrm{life}} =
\frac{51\,000\,\mathrm{y}}{B_{-9,\mathrm{IC}}(z)^2 +
B_{-9,\mathrm{jet}}^2} \left(\frac{B_{-9,\mathrm{jet}}}{\nu_{15}}
\right)^{\frac{1}{2}}, \label{eq:maxage} \end{equation} where
$B_{-9,\mathrm{jet}}$ is the magnetic flux density in nT of the
jet field, the background radiation energy density has been
expressed in terms of an equivalent magnetic field,
$B_{-9,\mathrm{IC}} = (1+z)^2\times 0.45\,\mathrm{nT}$, and
$\nu_{15}$ is the observing frequency in $10^{15}$\,Hz
\citep{vdLP69}.  Note that as the substituted $\gamma \propto
B^{-\frac{1}{2}}$, setting $B_{-9,\mathrm{jet}}=0$ is now meaningless.

To be fully adequate for electrons in a relativistic jet at
cosmological distances, the equation needs to be modified.  Firstly,
the frequency local to the source is $(1+z)$ times the observing
frequency because of the cosmological redshift.  Furthermore, the
radiating electron may be embedded in a relativistic flow with bulk
Lorentz factor $\Gamma$ with three consequences: relativistic time
dilation and Doppler shift, and boost of the background radiation
energy density.  The relativistic time dilation enhances the electron
lifetime in the jet frame by a factor $\Gamma$.  The Doppler shift
between the emission frequency $\nu_{\mathrm{int}}$ in the jet frame
(equal to the characteristic frequency) and the observation frequency
$\nu_{\mathrm{obs}}$ is given by \begin{math}\nu_{\mathrm{int}} =
\nu_{\mathrm{obs}} {\cal D}^{-1},\end{math} where ${\cal
D}(\Gamma,\theta) = [\Gamma(1-\beta_{\mathrm{jet}} \cos
\theta)]^{-1}$, the Doppler boosting factor for an angle $\theta$ to
the line of sight \citep[see ][ \emph{e.\,g.}]{HM91}.  A relativistic
flow perceives the energy density of the background radiation field
boosted up by a factor $\Gamma^2$ \citep{Der95}. Inserting these
yields \begin{equation} t_{\mathrm{life}}({\cal D},z) = \frac{\Gamma {\cal
D}^{\frac{1}{2}} \times 51\,000\,\mathrm{y}}{\left(\Gamma
B_{-9,\mathrm{IC}}\right)^2 + B_{-9,\mathrm{jet}}^2}
\left(\frac{B_{-9,\mathrm{jet}}}{(1+z)\nu_{15}}
\right)^{\frac{1}{2}}. \label{eq:beammax} \end{equation}

\begin{figure}
  \center\resizebox{0.8\hsize}{!}{\includegraphics{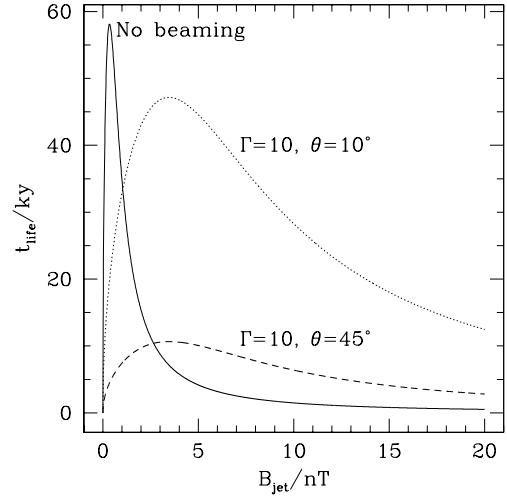}} \caption{Maximum
  age against synchrotron and inverse Compton cooling (off microwave
  background photons at 3C\,273's redshift) of an electron radiating at observed UV
  wavelengths plotted against the jet magnetic field.
  Solid line, no beaming; dashed line, relativistic
  Doppler boosting with $\Gamma=10$ and a line-of-sight
  angle $\theta=45\degr$. \label{f:maxage}}
\end{figure}

The electron lifetime attains a maximum value at a certain value of
the jet's magnetic field (Fig.~\ref{f:maxage}) \citep{vdLP69}.  On
either side of the maximum, the lifetime is decreased by larger losses
suffered \emph{before} it is observed: for a higher magnetic field in
the jet, the synchrotron cooling is more rapid ($t \propto
1/U_\mathrm{tot}$, Eqn.~\ref{eq:tofE}). A lower jet field requires an
electron of higher Lorentz factor for emission at the given frequency,
which also suffers more rapid losses ($t \propto 1/\gamma$,
Eqn.~\ref{eq:tofE}).  By differentiation of Eqn.~\ref{eq:beammax}, the
value of the \emph{maximum} lifetime is \begin{equation}
t_{\mathrm{max}}(B_{\mathrm{IC}}, {\cal D}, z) = \sqrt{\frac{\cal
D}{\Gamma}} \frac{29\,000\,\mathrm{y}}
{B_{-9,\mathrm{IC}}^{\frac{3}{2}} \left[(1+z)
\nu_{15}\right]^{\frac{1}{2}} }. \label{eq:beamabsmax} \end{equation} Note that
the largest possible value for the factor $\sqrt{{\cal D}/\Gamma}$ is
$\sqrt{2}$.  $t_{\mathrm{max}}$ is a firm upper limit for the lifetime
of a synchrotron-radiating electron from a source at redshift $z$ in a
flow of bulk Lorentz factor $\Gamma$, whatever the magnetic field
strength in the source. It is deduced only from the fact that
synchrotron emission is observed at a certain frequency, and that
electrons which can radiate at this frequency suffer drastic energy
losses either by synchrotron or by inverse Compton cooling between the
time of acceleration and the time of emission.  The only additional
assumptions are rapid pitch-angle scattering and a homogeneity of
conditions throughout the electron's lifetime.

The VLBI jet close to the core has a line-of-sight angle near
10\degr\ and a bulk Lorentz factor near 10 \citep{AR99}.  A
line-of-sight angle $\theta \approx 45\degr$ has been inferred for
the flow into the hot spot from independent considerations of the
jet's polarisation change there and the hot spot's morphology
\citep{jetIII,hs_II}.  We have plotted the lifetime of an electron
responsible for emission from the jet in 3C\,273 observed at
300\,nm as function of the jet's magnetic field in
Fig.~\ref{f:maxage} for the extreme cases of no beaming in the
optical jet and beaming identical to that in the VLBI jet with
$\Gamma=10, \theta = 10\degr$, and for an intermediate case with
$\Gamma=10, \theta=45\degr$ (though note that $\Gamma=10, \theta =
10\degr$ is unrealistic as there is a difference in position angle
between the VLBI jet at 244\degr\ and the arcsecond jet at
222\degr).

The equipartition flux densities derived for the jet lie in the
range of 15\,nT \citep{NeumDiss} up to 67\,nT for region\,A
\citep{Roe00}, leading to maximum ages of 100--800 years, less
than the light travel time from one bright region to the next.  In
the absence of beaming effects, the largest possible lifetime for
electrons in 3C\,273 from Eqn.~\ref{eq:beamabsmax} is about
58\,000\,y, again short of the required values.  The ``boosted
lifetime'' can be at most $\sqrt{2}$ larger than this.  There is
thus no combination of $\Gamma, \theta$ which enhances the
electron lifetime to the 100\,000\,y required for illumination of
the entire jet in 3C~273 by UV-radiating electrons.

Thus, the invocation of mild or even drastic beaming and/or
sub-equipartition fields cannot resolve the discrepancy between the
synchrotron loss scale and the extent of the optical jet of 3C\,273,
as has been possible for the jet in M87 \citep{HB97}.  As another
alternative to invoking quasi-permanent re-acceleration, the existence
of a ``loss-free channel'' in which electrons can travel down a jet
without synchrotron cooling has been proposed by \citet{Oweeta89}.  As
an extreme version of this case, we assume that the electron travels
along the jet in zero magnetic field and is observed as soon as it
enters a filament with magnetic field $B_{\mathrm{fil}}$.  The energy
loss between acceleration and synchrotron emission is then only due to
inverse Compton scattering. The lifetime in 3C\,273 is then
130\,000\,y $\times \Gamma^{-1} \sqrt{B_{\mathrm{-9,fil}} {\cal
D}/\nu_{15}}$. Again, if the jet flow in 3C\,273 is highly
relativistic, the electrons suffer heavy inverse Compton losses and
the lifetime mismatch persists.  In any case, it remains to be shown
that the ``loss-free channel'' is a physically feasible configuration
of an MHD jet.

\section{Conclusion}
\label{s:conc} We have analysed deep HST images at 300\,nm and
600\,nm (Figs. \ref{f:R-band} and \ref{f:U-band}) of the jet in
3C\,273 and constructed an optical spectral index map at 0\farcs2
resolution (Fig.~\ref{f:alpha}). The optical spectral index varies smoothly
over the entire jet, indicating a smooth variation of the physical
conditions across the jet. Unlike in M87 \citep{MNR96,Per01}, there is
no strong correlation between optical brightness and spectral index
(Figs. \ref{f:alpha-trace} and \ref{f:correlation}).  The spectral
index map thus shows no signs of strong synchrotron cooling at any
location in the jet.  Particle acceleration at a few localised sites
in the jet is not sufficient to explain the absence of strong cooling.
This does not preclude the possibility that the enhanced-brightness
regions are shocks -- but even if they are, re-acceleration between
them is necessary to explain the observed spectral index features.  We
have further shown that relativistic effects cannot lead to
significant enhancements of the electron lifetime in 3C\,273's jet,
whatever the bulk Lorentz factor (Sect.~\ref{s:disc.jet.cooling}),
strengthening previous electron lifetime arguments.  The need for a
continuous re-acceleration of electrons emitting high-frequency
synchrotron radiation in the jet of 3C\,273 is thus evident.

Mechanisms have been proposed which can explain the apparent lack of
cooling by distributed re-acceleration.  These include acceleration by
reconnection in thin filaments \citep{LB98} and turbulent acceleration
\citep{MAV99}.  Both processes manage to maintain the injection
spectrum over distances much larger than the loss scales, although the
latter so far only maintains cutoff frequencies in the range of
$10^{12}\,$Hz--$10^{13}\,$Hz, \emph{i.\,e.,} below the values
observed in 3C\,273.

We note that for those jets which \emph{are} bulk relativistic flows
at high Lorentz factors, the increased inverse Compton losses form a
further sink of energy that has to be filled by re-energization
processes inside the jets.  This requirement becomes more severe at
higher redshifts.  As has been suggested previously \citep{Cel00},
inverse Compton scattering off cosmic microwave background photons
might explain the so far unaccounted-for X-ray flux from 3C\,273's jet
\citep{Roe00,Mareta01} and from Pictor~A's jet and hot spot
\citep{Wil01}. If this is true, the outward-decreasing X-ray flux from
the jet in 3C\,273 indicates that the jet is still highly relativistic
near region~A and slows down towards the hot spot.  Recently, the
detection of extended X-ray emission from the jet in PKS~0637$-752$ by
the new X-ray observatory {\sc Chandra} has been reported and a
similar explanation has been brought forward
\citep{Cha00,Sch00,Tav00,Cel00}. Like for 3C\,273, \emph{in situ}
re-acceleration is required in PKS~0637$-752$ to explain the mismatch
between de-projected extent of the jet ($>1$\,Mpc) and inverse-Compton
loss scale (10\,kpc) \citep{Tav00}.

In order to understand the physical conditions in extragalactic jets
like that in 3C\,273, it is necessary to detect even the most subtle
variations in the parameters describing the synchrotron spectrum (that
is, cutoff frequency, break frequency and especially low-frequency
spectral index), which requires the deepest images at the highest
resolutions and in many wavelength bands to detect variations at all.
A theory of the physical processes at work in this jet initially has
to explain both the simple overall spectral shape, as well as its
constancy over scales of many kiloparsec.  The details of this
physical process will be constrained by the subtle deviations from the
simple spectral shape, such as those tentatively identified in
Sect.~\ref{s:disc.jet.alpha}.  We aim to find these deviations with the
full data set, including new radio and near-infrared data in addition
to the presented optical and UV images.

\begin{acknowledgements}
\label{s:ack}
We are grateful to D.~van Orsow for his
assistance with the HST observations.  This research has made use of
NASA's Astrophysics Data System Abstract Service.
\end{acknowledgements}

\appendix
\section{Image alignment}
\label{a:alignment}

\subsection{Flux errors from misalignment}
\label{s:flux_error}

Consider two images which are not registered correctly and which have
slightly different PSF widths.  When performing photometry on these
images, we assume that they are registered perfectly and have
identical, known beam sizes.  This amounts to making a flux
measurement in a certain aperture in one image, but in a slightly
offset aperture of slightly different size in the second image.  The
error is largest in the steepest gradients in the image, which are the
flanks of the PSF of width $\sigma$, smoothed to the desired effective
beam size $\sigma_\mathrm{eff}$, with $\sigma_\mathrm{eff}^2 =
\sigma^2 + \sigma_\mathrm{smooth}^2$.  Its magnitude can be assessed
by considering the PSF as a Gaussian at given position and of given
width, and as smoothing filter a second Gaussian slightly displaced
from the PSF and of width slightly different from that achieving the
desired effective beam area.  The result of the performed \emph{wrong}
flux measurement is then proportional to the integral
\begin{displaymath}
\int\int_{-\infty}^{\infty} \exp \left(-\frac{x^2 +
y^2}{2 \sigma^2_{\mathrm{PSF}}} \right) \exp \left(-\frac{(x-\delta x)^2 +
y^2}{2(\sigma_{\mathrm{smooth}} + \delta \sigma)^2} \right)
\mathrm{d}x \mathrm{d}y,
\end{displaymath}
where $\delta x$ is the offset of the aperture from the correct
position and $\delta \sigma$ is the error in the determination of the
PSF width.  The correct measurement is obtained by setting $\delta
x=0$ and $\delta \sigma=0$, and from this one obtains the fractional
flux error as a function of the two errors.  The fractional flux error
$\Delta f$ for a misalignment is
\begin{eqnarray}
\Delta f & = & 1 - e^{\frac{\delta x^2}{2\sigma_\mathrm{smooth}^2}} \nonumber \\
    &\approx & \frac{\delta x^2}{2\sigma_\mathrm{smooth}^2} \mathrm{\ if\ }
\frac{\delta x^2}{2\sigma_\mathrm{smooth}^2} \ll 1 \\
\label{eq:shift_error}
    & \Leftrightarrow & \frac{\delta x}{\mathrm{FWHM_\mathit{eff}}} =
    \sqrt{\frac{\Delta f}{4 \ln(2)}} \approx 0.6 \sqrt{\Delta f}.
\label{eq:shift_error_limit}
\end{eqnarray}
Hence, the relative flux error is better than 5\% if the misalignment
$\delta x < 10\%$. Similarly, for a wrong PSF width, the fractional error is
\begin{equation}
\Delta f = \frac{2\delta\sigma}{\sigma_\mathrm{smooth}}
\frac{\sigma^2}{\sigma^2_\mathrm{eff}} \mathrm{\ if\ }
\delta\sigma \ll \sigma_\mathrm{smooth}.
\end{equation}
This error is negligible if the desired effective PSF is much larger
than the intrinsic PSF of the input images, as is the case here.

\subsection{Refined image alignment}
\label{s:alignment}
\begin{figure}
  \resizebox{\hsize}{!}{\includegraphics{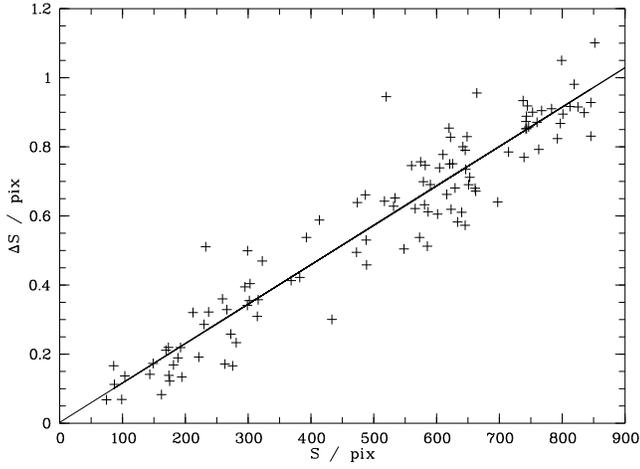}}
	\caption{The
  separation of a number of stars was measured both on a F336W (UV)
  and a F673W (red) image of R136. This plot shows the difference in
  separation between the frames, $\Delta S$, plotted against the
  separation on the ``red'' image, $S$. The best-fitting straight line
  is shown. $\Delta S$ grows systematically with $S$, indicating a
  differing scale between the frames. The slope of the line is
  $(1.14\pm .04)\times 10^{-3} $, the intercept with the ordinate is
  $(0.3 \pm 8)\times 10^{-2}$.  The slope is not changed within the
  errors by forcing the line to pass through the origin.}
  \label{f:r136}
\end{figure}

The two obvious ways to determine the relative shift between any two
images are measuring the positions of point sources on the various
images, or using the engineering files (also termed \emph{jitter
files}) provided as part of the observing data package.  If the
telescope pointing was known to be precise to better than 0\farcs02,
we could simply rely on the commanded shifts.

The HST has a pointing repeatability \emph{within a single
telescope visit\footnote{A \emph{visit} is a set of exposures that
should be observed together; the HST equivalent of an observing
run.  The telescope is aligned to its guide stars only at the
beginning of a visit.}} of about 5~milli-arcsecond (mas). The
offsets between individual exposures are accurate to 15~mas,
leading to a total error of about 16~mas. This is already
comparable to the demanded accuracy. The situation is expected to
be worse for relating exposures in different visits, when the
telescope has been pointing elsewhere in the mean time.

Unfortunately, measuring the positions of only a few (four, in our
case) astrometric reference stars does not immediately lead to accurate
measurements of the telescope's pointing -- especially since in our
case, each star lies on a different chip. The undersampling of the
telescope PSF by the WFPC2 pixels and the so-called sub-pixel
scattering of the WFPC2 detectors lead to an additional scatter of the
centroid positions of point sources, approximately uniformly
distributed between $+0.25$ and $-0.25$ pixels and \emph{in excess of
statistical uncertainties} (Lallo, 1998, \emph{priv. comm.}). With
only a small number of centroidable point sources available, the
centroiding errors are of the same magnitude as the intrinsic pointing
errors of the telescope.

It is therefore worth considering the pointing error sources
\emph{en d\'etail} to ensure that the alignment is at the required
0\farcs02 level\footnote{The use of the quasar's diffraction spike
by \citet{Baheta95} to locate the QSO's center is only expected to
work well for sources near the camera center (Krist, \emph{priv.
comm.}). We failed to reproduce their quoted accuracy with our
data.}. Alignment errors can be caused by roll or pointing errors
and less obviously by a scale difference between exposures using
different filters in the same camera. The importance of the
various alignment error sources can be estimated by considering
the effect they have on the hot spot location if the quasar images
are assumed to coincide. The hot spot is separated by 22\arcsec\
from the quasar, corresponding to about 480 PC pixels.

The roll repeatability of the HST is at the 10\arcsec\ level.  The
engineering data provided with HST exposures record roll angle
differences of about 6\arcsec. This is well below the rotation of
3\farcm5 which would produce a 0.44 pixel difference over 480 pixels.
The telescope roll differences can thus be neglected.

The engineering files record the \emph{telescope} pointing in
three-second intervals and can be used to calculate the offsets.
Their accuracy is only limited by the so-called ``jitter'', vibrations
due to thermal effects.  The jitter was below 10~mas in all exposures,
and below 5~mas in most.  There is an additional uncertainty from the
transformation between the telescope's focal plane and the detector:
the location of a camera inside the telescope may change slightly over time
(shifts, rotations, or both).  This uncertainty is irrelevant for
relative positions as long as the location and orientation of WFPC2
and the Fine Guidance Sensors (FGS, these perform the guiding
observations) inside the telescope is stable, which is the case for
the employed shifts of about 1\arcsec\ and for the timescales between
the visits.

Within a single visit, the engineering file information is used to
obtain relative offsets, with a typical 5~mas error.  The values
differ from the commanded shifts by a few milli-arcseconds at most.
All of these offsets are by an integer number of PC pixels.

Between the various visits, the telescope has been pointing to a
different part of the sky.  One should therefore not assume that the
relative shifts between various visits as determined from the jitter
files are as accurate as shifts within one visit.  We therefore
measured the positions of the four astrometric reference stars on each
of the short exposures (in fact, one of the stars is very faint in the
UV, so the position was determined on a long exposure for this one).
The shifts determined from the four point sources' positions have a
typical standard deviation (accuracy) of 15~mas in each coordinate.
On the three visits' sum images aligned this way, the scatter of the
quasar image position is less than 5~mas and 7~mas in x and y,
respectively.  This means that although the measured shifts have a
fairly large scatter, the resulting value is \emph{precise} to about
10~mas.  Finally, we note that the observed shifts of the stellar
positions and those obtained from the jitter files agree to better
than 0\farcs02 in all cases, with an RMS value of 0\farcs01.  There
are, however, systematic differences between these and the commanded
values.  Hence, we do not blindly rely on the latter.

Because of differential refraction in the MgF$_{2}$ field flattener
windows employed in WFPC2 \citep{trauger-geom}, the pixel scales of
images taken through the F622W and F300W filters differ by about
0.1\%. This alone is enough to eat up the alignment error budget of
0.44 pixels over 480 pixels separation.  The wavelength dependence was
expected from ray-tracing studies of the WFPC2 optics. Its presence
and magnitude were experimentally confirmed by comparing archival
images\footnote{Data from proposal 5589, PI John Trauger} of the star
cluster R136 taken through similar filters as those employed in the
present work (F336W and F672N) (Fig.\,\ref{f:r136}). The scale
difference has to be removed before combining the images to a spectral
index map. This was done in the following manner: the plate scale of
each image was calculated using the parameters in
\citet{trauger-geom}. All images were resampled to a grid with pixel
size of 0\farcs0045548, which is one tenth of the average of the
original scales, using bilinear interpolation. The result was then
binned in blocks of $10\times 10$ pixels to a common pixel size of
0\farcs045548.  The scale of the two images is then identical to
better than 1 part in 10\,000.

The resampling required the use of the quasar image as common
reference point between the two filters.  The QSO's position can be
determined to about 10--15~mas by centroiding routines on the unsaturated
images.

\end{document}